%% file: conf.tex
\documentclass[aps,prl,preprint,tightenlines,superscriptaddress,showpacs,byrevtex]{revtex4}
\usepackage{epsfig, amssymb}

\begin{document}
\advance\hoffset by  -4mm

\newcommand{\de}{\Delta E}
\newcommand{\mbc}{M_{\rm bc}}
\newcommand{\bb}{B{\bar B}}
\newcommand{\qq}{q{\bar q}}
\newcommand{\ks}{\bar{K_S}}
\newcommand{\kpi}{K^-\pi^+}
\newcommand{\kpipin}{\kpi\pi^0}
\newcommand{\kpipipi}{\kpi\pi^-\pi^+}
\newcommand{\kpipi}{K^-\pi^+\pi^+}
\newcommand{\kspi}{\ks\pi^+}
\newcommand{\dndp}{D^+ \bar{D}^0}
\newcommand{\dndm}{D^- D^0}
\newcommand{\dndb}{\bar{D}^0 D^0}
\newcommand{\bdndp}{B^+\to\dndp}
\newcommand{\bdndm}{B^-\to\dndm}
\newcommand{\bdndb}{B^0\to\dndb}
\newcommand{\brdndp}{(3.85\pm 0.31\pm 0.38)\times 10^{-4}}
\newcommand{\brdndb}{0.42\times 10^{-4}}
\newcommand{\acpnbdp}{0.00\pm 0.08 \pm 0.02}
\newcommand{\acpul}{-0.14<A_{CP}<0.14}
\newcommand{\br}{{\mathcal B}}

\preprint{\vbox{ \hbox{   }
    \hfill
    \hbox{BELLE-CONF-0762}
}}

\title{\Large \rm\quad\\[0.5cm] 
Measurement of $\bdndp$ branching fraction and charge 
asymmetry and search for $\bdndb$
}
\input {author-conf2007.tex}
\noaffiliation

\begin{abstract}
We report an improved measurement of the
$B^+\to D^+\bar{D}^0$ and $B^0\to D^0\bar{D}^0$
decays based on $656.7\times 10^6$ $\bb$ events
collected with the Belle detector at KEKB.
We measure the branching fraction and charge asymmetry for the $\bdndp$
decay: $\br(\bdndp)=\brdndp$ and $A_{CP}(\bdndp)=\acpnbdp$,
where the first error is statistical and the second is systematical.
We also set the upper limit for the $\bdndb$ decay:
$\br(\bdndb)<\brdndb$ at 90\% CL.
\end{abstract}
\maketitle

\section{Introduction}
Recently, evidence of direct $CP$ violation in $B^0\to D^+ D^-$ decays
was observed by the Belle collaboration~\cite{belle_dpdm}. 
A possible interpretation is that there is a sizable penguin contribution
to this decay. If this is the case,
a similar effect should be seen in the charged mode $\bdndp$, and
has been already observed by Belle~\cite{belle_dndp} and confirmed by 
BaBar~\cite{babar_dndp}.
The inclusion of
charge conjugate states is implicit throughout this paper.
In this paper, we report an improved measurement of
the branching fraction and charge asymmetry for $\bdndp$ decay and
we also search for the $\bdndb$ decay.
The latter can only be produced by $W$ exchange diagram.
We use a data sample of $656.7\times 10^6$ $\bb$ events
collected with the Belle detector at the KEKB collider~\cite{KEKB}.

\section{Belle detector}

The Belle detector is a large-solid-angle magnetic
spectrometer that consists of a silicon vertex detector (SVD),
a 50-layer central drift chamber (CDC), an array of
aerogel threshold Cherenkov counters (ACC),
a barrel-like arrangement of time-of-flight
scintillation counters (TOF), and an electromagnetic calorimeter (ECL)
comprised of CsI(Tl) crystals located inside
a superconducting solenoid coil that provides a 1.5~T
magnetic field.  An iron flux-return located outside of
the coil is instrumented to detect $K_L^0$ mesons and to identify
muons (KLM).  The detector is described in detail elsewhere~\cite{NIM}.
For the first sample
of 152 million $B\bar{B}$ pairs, a 2.0 cm radius beam pipe
and a 3-layer silicon vertex detector were used;
for the latter 505 million $B\bar{B}$ pairs,
a 1.5 cm radius beam pipe, a 4-layer silicon detector
and a small-cell inner drift chamber were used~\cite{Ushiroda}.

\section{Events selection}
The track transverse momentum is required to be higher than 
0.075~GeV$/c$ in order to reduce combinatorial background. 

For charged particle identification (PID), 
the measurement of the specific ionization ($dE/dx$) in the CDC,
and signals from the TOF and by ACC are used.
Charged kaons are selected with PID criteria that have
an efficiency of 88\% with a pion misidentification probability of 8\%.
All charged tracks that are consistent with a pion
hypothesis that are not positively identified as electrons are 
treated as pion candidates.

Neutral kaons are reconstructed in the decay $K_S\to\pi^+\pi^-$;
no PID requirements are applied for the daughter pions.
The two-pion invariant mass is required to be within 9~MeV$/c^2$
($\sim 3\sigma$) of the $K^0$ mass and the displacement of the 
$\pi^+\pi^-$ vertex from the IP in the transverse ($r-\varphi$) plane is 
required to be between 0.2~cm and 20~cm. 
The $K_S$ momentum and the vector from the IP to the
$\pi^+\pi^-$ vertex are required to be collinear in the $r-\varphi$ 
plane to within 0.2 radians.

Photon candidates are selected from ECL showers not associated
with charged tracks.
An energy deposition of at least 75~MeV and a photon-like shape 
of the shower are required for each candidate.
A pair of photons with an invariant mass 
within 12~MeV$/c^2$ ($\sim 2.5\sigma$) of the $\pi^0$ mass is 
considered as a $\pi^0$ candidate.

We reconstruct $\bar{D}^0$ mesons in the $\kpi$, $\kpipipi$ and 
$\kpipin$ decay channels. The $D^+$ candidates are reconstructed in
$K^-\pi^+\pi^+$ and $\ks\pi^+$ final states.
We require the invariant mass of the $\bar{D}^0 (D^+)$ candidates to 
be within 11~MeV$/c^2$ ($1.5\sigma$ for $\kpipin$ and 
$2.5\sigma$ for other modes) of the nominal $\bar{D}^0 (D^+)$ mass.
For $\bar{D}^0\to\kpipin$ decay, we require that
the $\pi^0$ momentum be greater than 0.35~GeV$/c$ in order to reduce 
further the combinatorial background.
To suppress the large background from $B^+\to D_s^+\bar{D}^0$ decays
with the $K^+$ from $D_s^+$ misreconstructed as a pion, none of the
pions from $D^+$ should be consistent with the kaon hypothesis.
This requirement has an efficiency of 93\% and kaon misidentification 
probability of 9\%.

We combine $\bar{D}^0$ and $D^+$ ($D^0$) candidates to form $B^+$ ($B^0$) 
candidates.
These are identified by their center-of-mass (CM) energy difference, 
\mbox{$\de=(\sum_iE_i)-E_{\rm beam}$}, and the beam constrained mass, 
$\mbc=\sqrt{E_{\rm beam}^2-(\sum_i\vec{p}_i)^2}$, where $E_{\rm beam}$ 
is the beam energy and $\vec{p}_i$ and $E_i$ are the momenta and 
energies of the decay products of the $B$ meson in the CM frame. 
We select events with $\mbc>5.2$~GeV$/c^2$
and $|\de|<0.3$~GeV, and define a $B$ signal region of $|\de|<0.02$~GeV,
$5.273$~GeV$/c^2<\mbc<5.287$~GeV$/c^2$.
In an event with more than one $B$ candidate, we choose the one
with smallest $\chi^2$ from $D$ mass fit.
We use a Monte Carlo (MC) simulation to model the response of
the detector and determine the efficiency~\cite{GEANT}.

Variables that characterize the event topology are used to suppress 
background from the jet-like $e^+e^-\to\qq$ continuum process.
We require $|\cos\theta_{\rm thr}|<0.8$, where $\theta_{\rm thr}$ is 
the angle between the thrust axis of the $B$ candidate and that of the 
rest of the event; this condition rejects 77\% of the continuum 
background while retaining 78\% of the signal. 
To suppress high background in the $\dndb$ final state, we use a
Fisher discriminant, ${\cal F}$, that is based on the production
angle of the $B$ candidate, the angle of the $B$ candidate thrust axis
with respect to the beam axis, and nine parameters that characterize
the momentum flow in the event relative to the $B$ candidate thrust
axis in the CM frame~\cite{VCal}. We impose a requirement on
${\cal{F}}$ that rejects 52\% of the remaining continuum background
and retains 86\% of the signal.

\section{Results}
\subsection{Measurement of the branching fractions}

\begin{figure*}
  \includegraphics[width=0.42\textwidth] {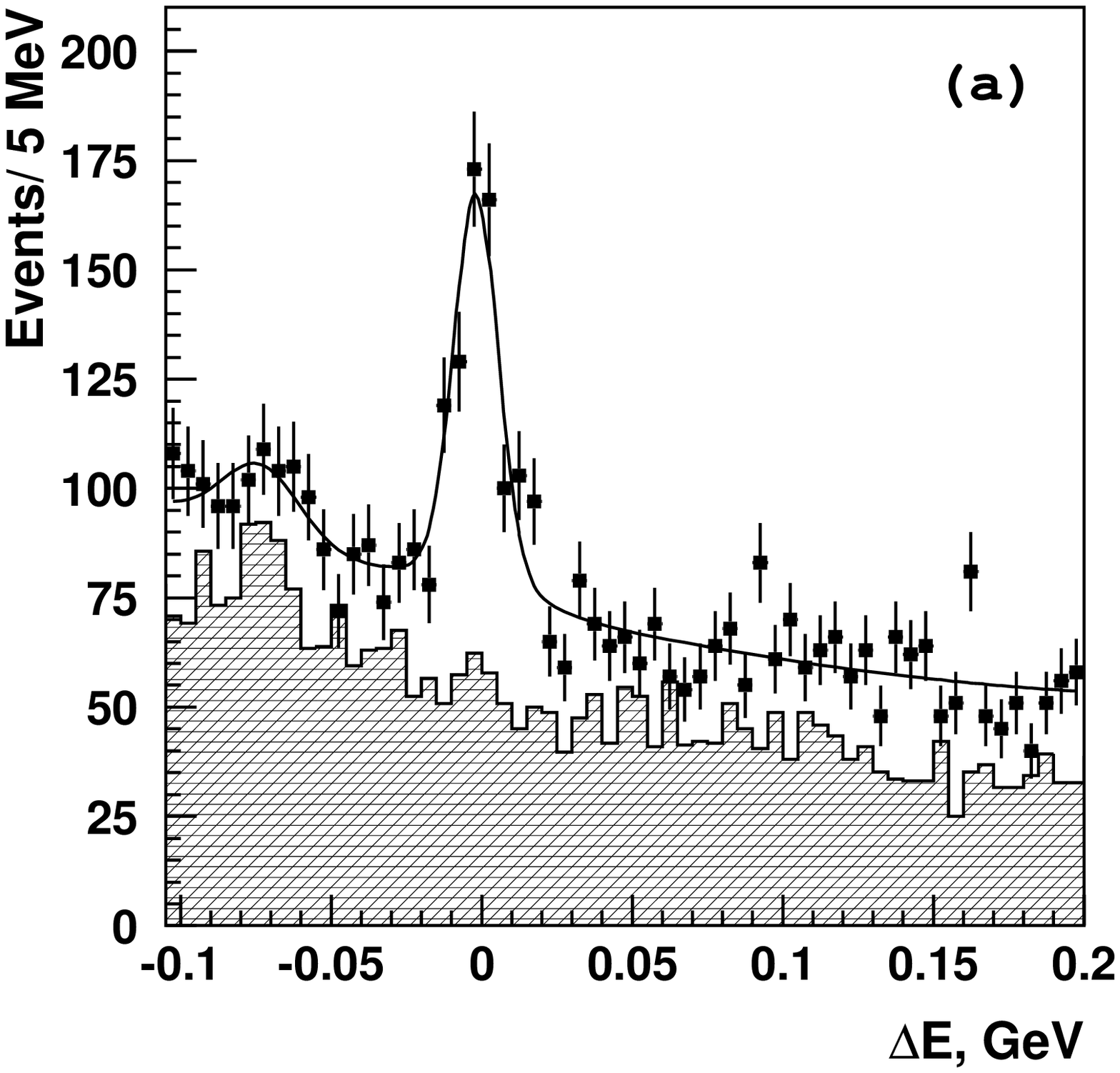} \hfill
  \includegraphics[width=0.42\textwidth] {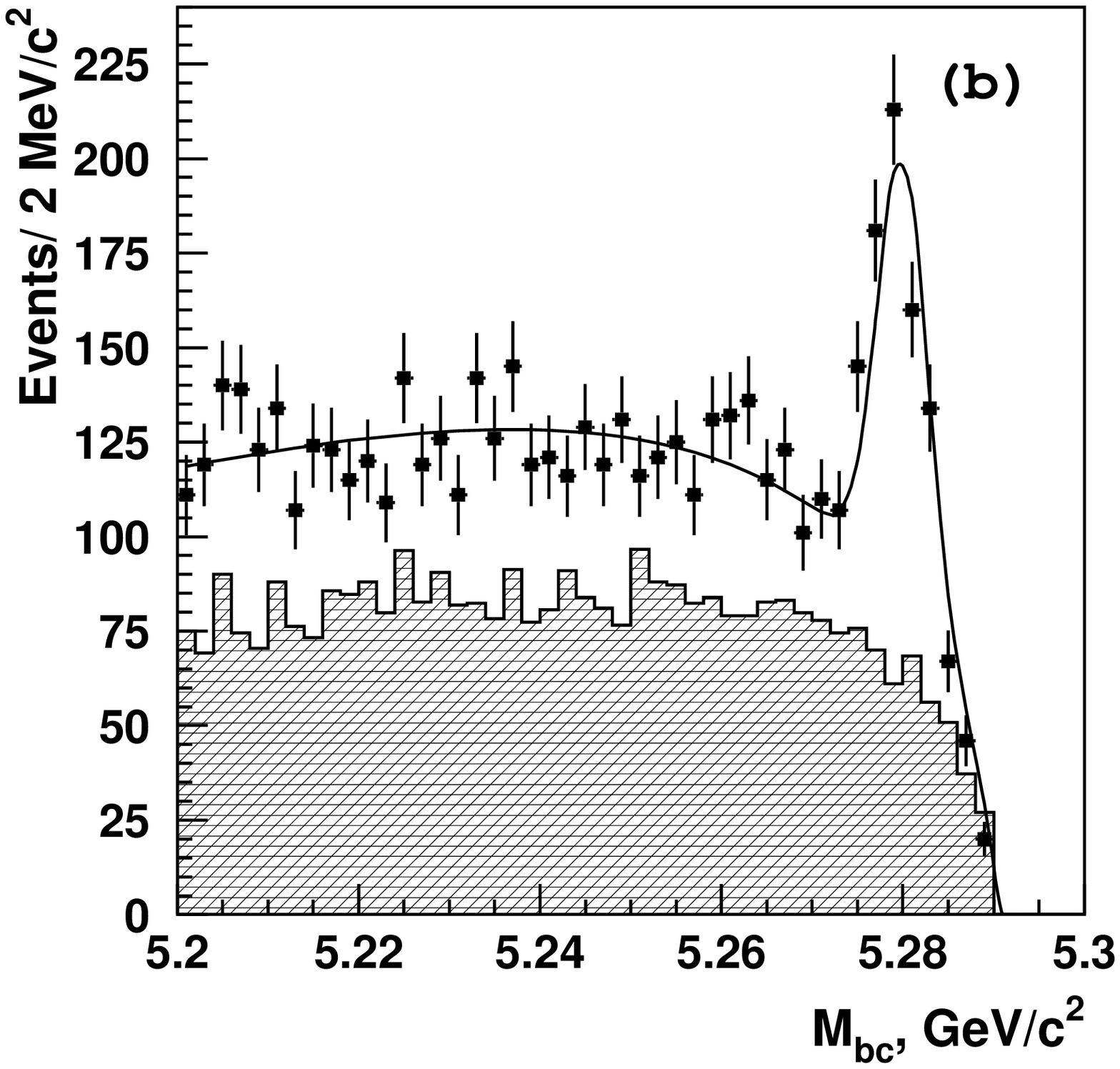}
  \caption{$\de$ (a) and $\mbc$ (b) distributions for the
    $\bdndp$ candidates. Each distribution is the projection of the 
    signal region of the other parameter. Points with errors
    represent the experimental data, crosshatched histograms show
    the $\bb$ MC and curves are projections from the two dimensional fits.}
  \label{dndc_mbcde}
\end{figure*}

\begin{figure*}
  \includegraphics[width=0.42\textwidth] {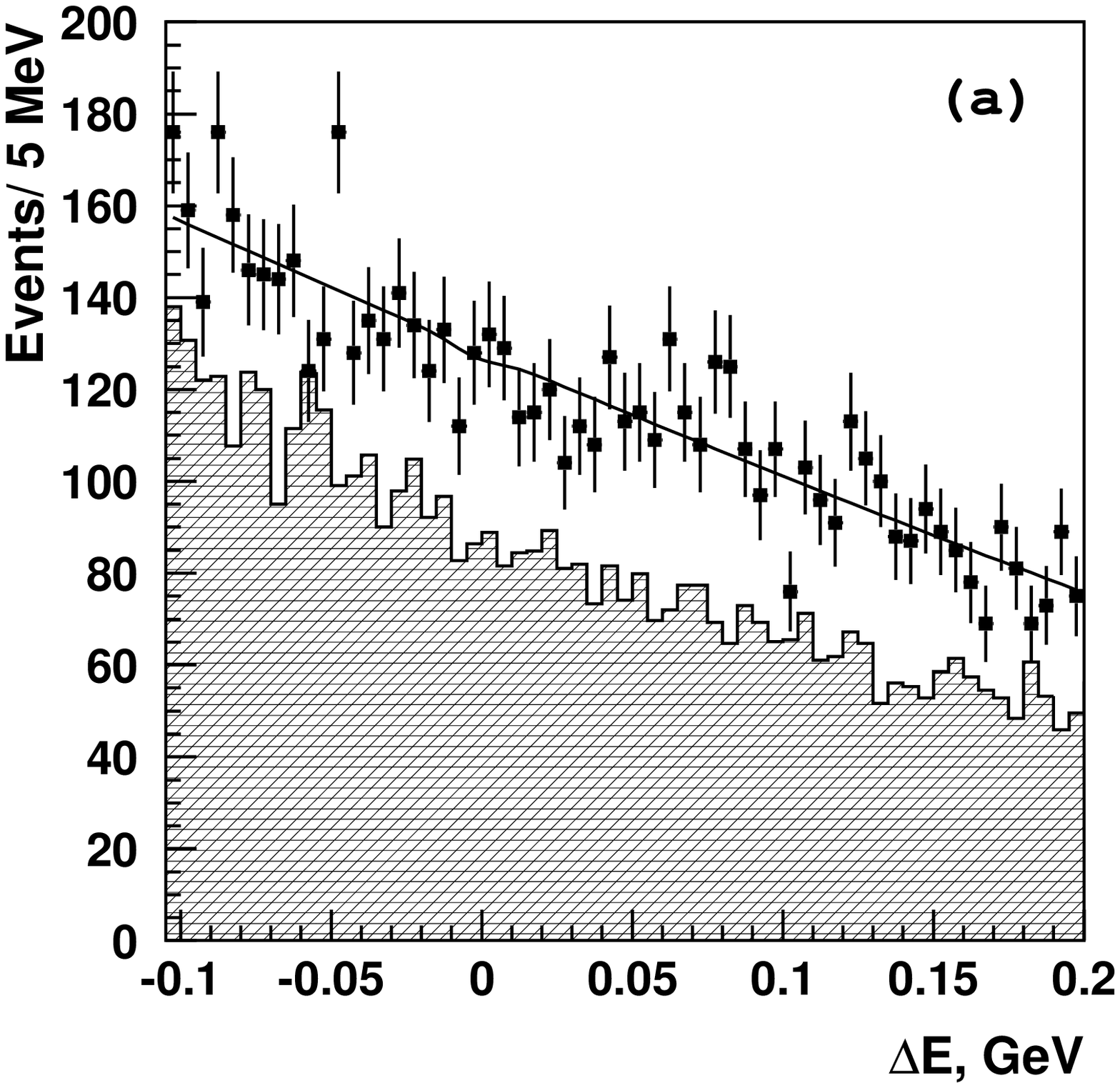} \hfill
  \includegraphics[width=0.42\textwidth] {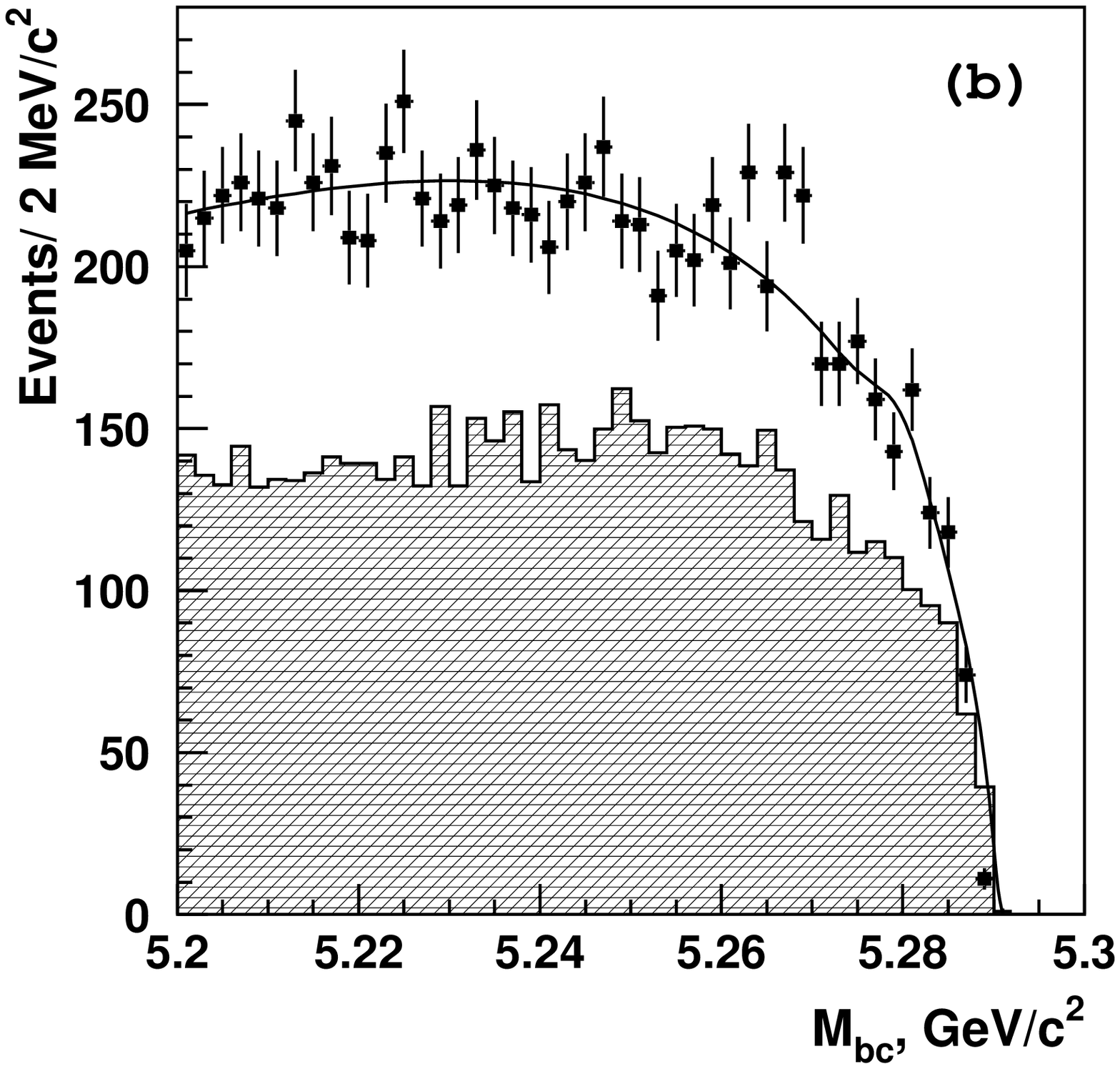}
  \caption{$\de$ (a) and $\mbc$ (b) distributions for the
    $\bdndb$ candidates. Each distribution is the projection of the 
    signal region of the other parameter. Points with errors
    represent the experimental data, crosshatched histograms show
    the $\bb$ MC and curves are projections from the two dimensional fits.}
  \label{dndb_mbcde}
\end{figure*}

The $\de$ and $\mbc$ distributions for 
$\bdndp$ and $\bdndb$ candidates are presented in 
Figs.~\ref{dndc_mbcde} and \ref{dndb_mbcde}.
The region $\de<-0.1$~GeV is excluded from the fit 
to avoid contributions from $B\to\bar{D} D^*$ decays.

The signal probability density function (PDF) is described by double 
Gaussian for $\de$ and a single 
Gaussian for $\mbc$. The $\de$-$\mbc$ correlation is taken into account.
We use the $B^+\to D_s^+\bar{D}^0$ events in our data sample to calibrate 
the means and resolutions of the signal shape.
The continuum, $\bb$ and $B\to D_s D^0$ background contributions are
described separately.
We use a linear function for $\de$ and threshold function for $\mbc$ 
to describe the continuum PDF.
The $\bb$ background is modeled by a quadratic polynomial for $\de$, 
a threshold function for $\mbc$ combined with a small peaking component 
(wide Gaussian on $\de$, Gaussian for $\mbc$). 
The shape of peaking background and threshold function 
parameters are fixed from the generic $\bb$ MC.
The $\de$ linear slope and quadratic term are free parameters.
The peak in the $\de$ distribution near $-70$ MeV coming from the 
$B\to D_s D^0$ decay is described by
a Gaussian for $\de$ and a Gaussian for $\mbc$.
Again, we use $B^+ \to D_s^+ \bar{D}^0$ to obtain the
parameters of this PDF.

\begin{table*}
\caption{Yields from the $\de$, $\mbc$ and 2D ($\de$-$\mbc$) fits, 
detection efficiencies, and corresponding branching fractions.
Errors are statistical only.  Upper limits are at the 90\% confidence level.
}
\medskip
\label{defit}
\begin{tabular*}{\textwidth}{l@{\extracolsep{\fill}}ccccc}\hline\hline
Decay channel & $\de$ yield & $\mbc$ yield & 2D yield & $\varepsilon$, $10^{-4}$& 
${\cal B}$, $10^{-4}$\\\hline

$\bdndb$ & $-4.5\pm 29.7$ & $5.7\pm 28.6$ & $0.4\pm 24.8$ ($<41$)& 
$16.4$ & $<0.38$ \\

$B^\pm\to D^\pm D^0$ & $366.4\pm 31.8$ & $376.4\pm 30.7$ & $369.7\pm 29.4$ & 
$14.6$ & $3.85\pm 0.31$ \\
\hline\hline
\end{tabular*}
\end{table*}

We determine the signal yield from the two-dimensional (2D) fit to the 
$\de$-$\mbc$ distribution. As a cross-check, we also do separate
one-dimensional fits for the $\de$ and $\mbc$ distributions, 
with the value of the other variable being in the signal region. 
The results are given in Table~\ref{defit}, 
where the listed efficiencies include intermediate branching fractions.
The projections of the 2D fit result is shown in
Figs.~\ref{dndc_mbcde} and \ref{dndb_mbcde}.

\subsection{Charge asymmetry in $\bdndp$ decay}
\begin{table*}
\caption{Charged $B$ mesons yields from the $\de$, $\mbc$ and 
2D ($\de$-$\mbc$) fits.
Errors are statistical only.
}
\medskip
\label{asymfit}
\begin{tabular*}{\textwidth}{l@{\extracolsep{\fill}}ccc}\hline\hline
Decay channel & $\de$ yield & $\mbc$ yield & 2D yield\\\hline
$\bdndp$ & $183.9\pm 21.5$ & $184.4\pm 21.4$ & $184.2\pm 20.4$\\
$\bdndm$ & $183.4\pm 22.1$ & $192.5\pm 21.8$ & $185.4\pm 21.0$\\
\hline\hline
\end{tabular*}
\end{table*}

\begin{figure}
  \includegraphics[width=0.45\textwidth] {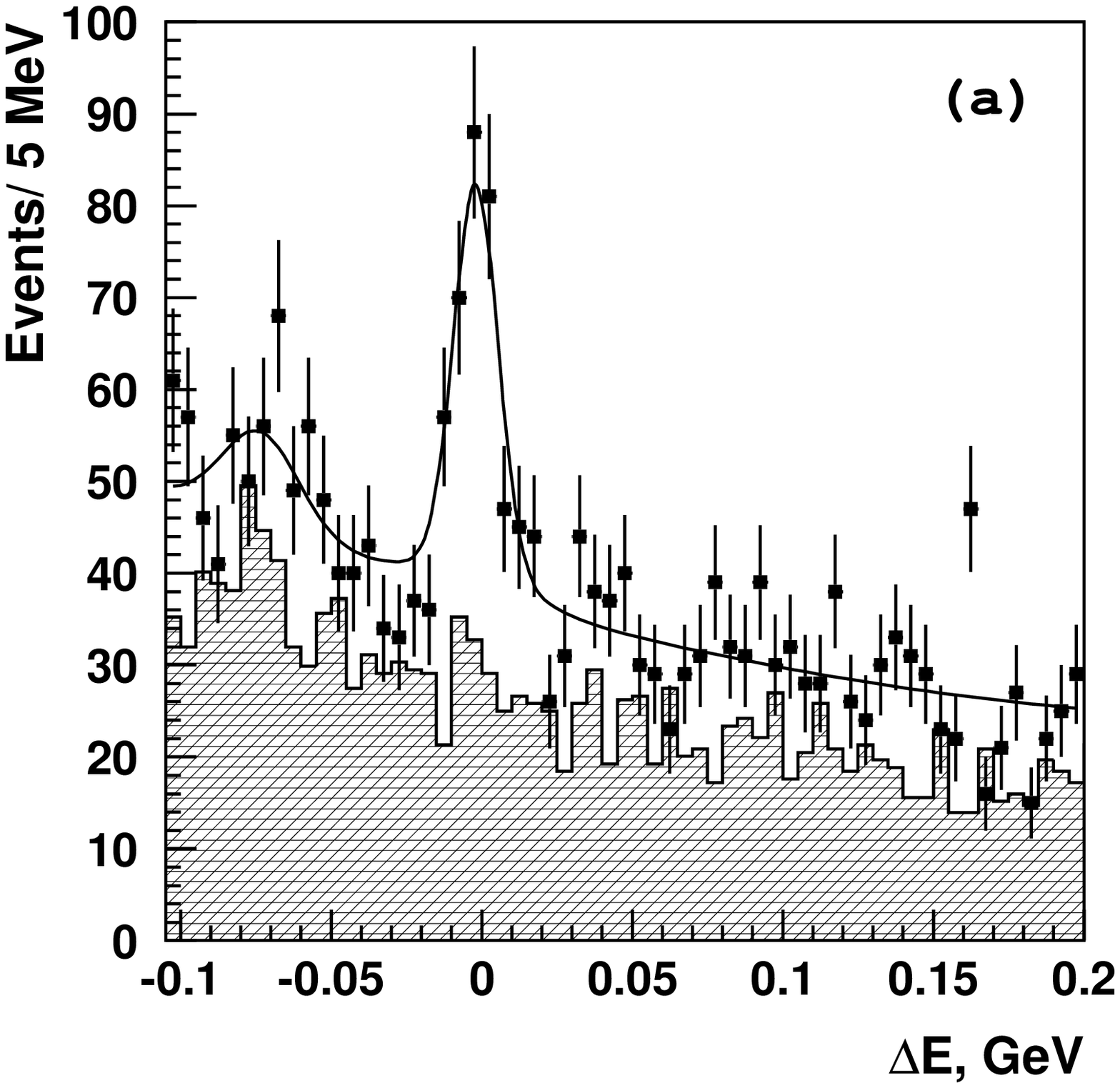}\hfill
  \includegraphics[width=0.45\textwidth] {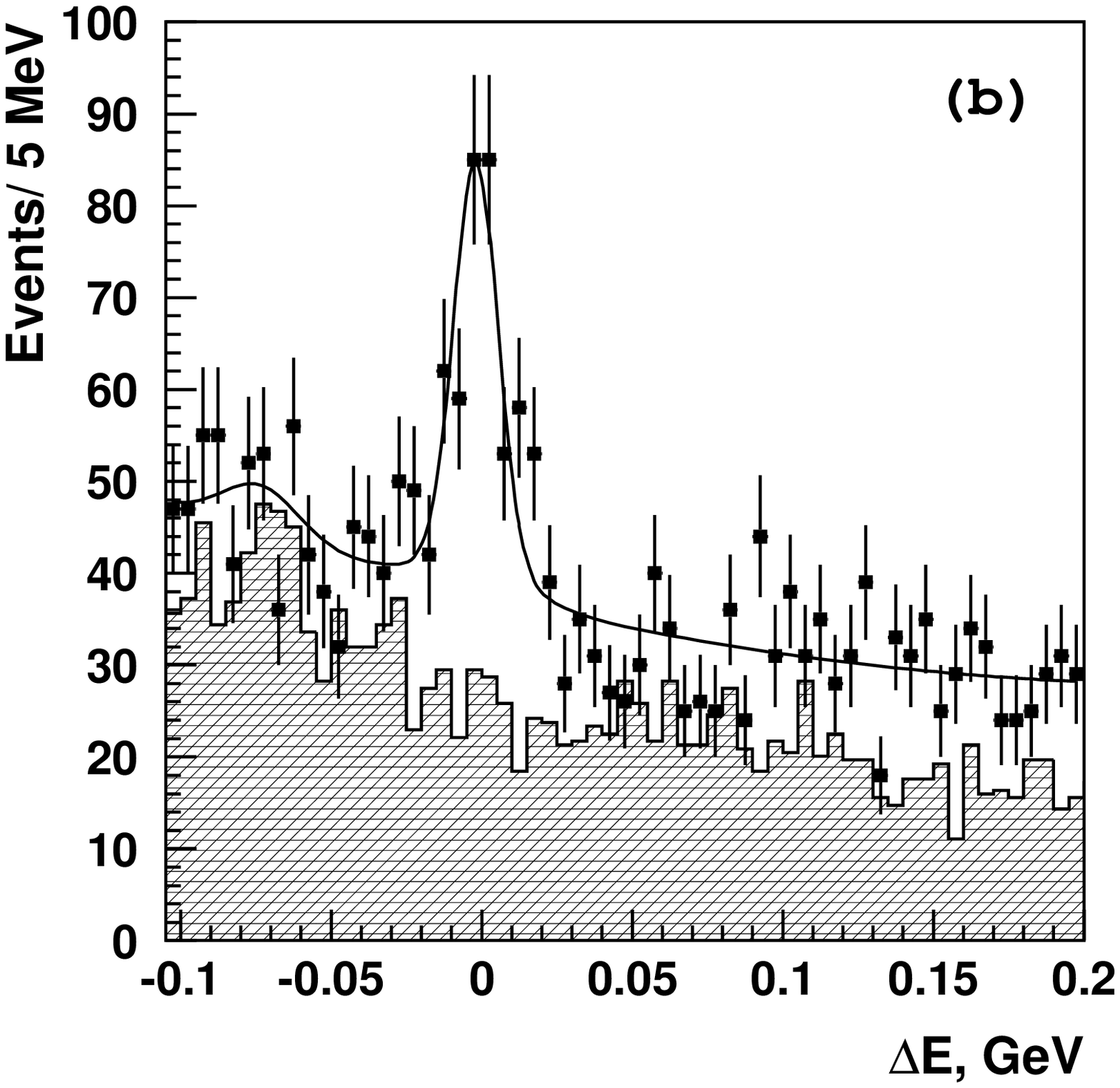}
  \caption{The $\de$ distributions for the $\bdndp$ (a)
    and the $\bdndm$ (b).
    The points with error bars show the data, the hatched histograms 
    represent $\bb$ MC and the curves are projections from the 2D fits.}
  \label{dndc_de}
\end{figure}

To calculate the charge asymmetry in the $\bdndp$ decay channel, we repeat 
the fits  separately for the $\bdndp$ and $\bdndm$ samples.
The $\de$ distributions for $\bdndp$ and $B^- \to D^- D^0$ candidates are 
presented in Fig.~\ref{dndc_de}.
The fit results are given in Table~\ref{asymfit}.
Using the results of the 2D fits, we calculate the charge asymmetry:
\begin{equation}
A_{CP}=\frac{N(\dndm)-N(\dndp)}{N(\dndm)+N(\dndm)}=0.00\pm 0.08
\end{equation}
where the error is statistical only.

\subsection{Cross checks \& systematic uncertainties}
We calculate the $\bdndp$ branching fraction separately for all used $D$ 
decay channels, and the results are consistent with the average value.
As an additional check, we apply a similar procedure to a decay chain 
with similar final state: $B^+\to D_s^+\bar{D}^0$.
We measure the branching fraction 
$\br(B^+\to D_s^+\bar{D}^0)=(0.95\pm 0.02)\%$,
where the error is statistical only. This is
consistent with the world average value $(1.09\pm 0.27)\%$~\cite{PDG}.
The charge asymmetry in this final state is consistent with zero: 
$(-0.5\pm 1.5)\%$.
We also measure the charge asymmetry for the $D^+\bar{D}^0$ background 
events and find a value consistent with 0: $(-1.4\pm 1.3)\%$.

The following sources of systematic errors are considered:
tracking efficiency (6\%), PID
efficiency (2\%),
$\pi^0$ reconstruction efficiency (6\%),
$D$ branching fraction uncertainties (6\%), 
signal yield determination (4\%), 
luminosity determination (1.5\%) and MC statistics (1\%).
The uncertainty in the tracking efficiency is estimated using 
partially reconstructed $D^{*+}\to D^0[K_S^0\pi^+\pi^-]\pi^+$ decays. 
The kaon identification uncertainty is determined from 
$D^{*+}\to D^0[K^-\pi^+]\pi^+$ decays.
The error in signal yield determination is estimated by varying the 
signal and backround shapes and fit range.
We assume equal production rates for $B^+B^-$ and $B^0\bar B^0$ pairs 
and do not include the uncertainty related to this assumption in the 
total systematic error. 
The overall systematic uncertainty is 10\% for the branching fraction 
measurement.
The systematic uncertainty is taken into account in the upper limit 
calculation: the signal efficiency is decreased by the systematic 
uncertainty.

The asymmetry measurement contains the following systematic errors: 
tracking efficiency difference for $\pi^\pm$ (0.013), 
pion identification efficiency difference for $\pi^\pm$ (0.004)
and signal yield determination (0.015). 
The total systematic uncertainty is 0.02.

\section{Conclusion}
In summary, we report improved measurements of 
$\bdndp$ branching fraction and the charge asymmetry:
$\br(\bdndp)=\brdndp$ and $A_{CP}(\bdndp)=\acpnbdp$.
We also set the upper limit for the $\bdndb$ decay branching fraction
of $\br(\bdndb)<\brdndb$ at 90\% CL.
These results are consistent with our previous
results~\cite{belle_dndp} and supersede them.
Our results are also consistent with BaBar measurements~\cite{babar_dndp}.

\section{Acknowledgments}
We thank the KEKB group for the excellent operation of the
accelerator, the KEK cryogenics group for the efficient
operation of the solenoid, and the KEK computer group and
the National Institute of Informatics for valuable computing
and Super-SINET network support. We acknowledge support from
the Ministry of Education, Culture, Sports, Science, and
Technology of Japan and the Japan Society for the Promotion
of Science; the Australian Research Council and the
Australian Department of Education, Science and Training;
the National Science Foundation of China and the Knowledge
Innovation Program of the Chinese Academy of Sciences under
contract No.~10575109 and IHEP-U-503; the Department of
Science and Technology of India; 
the BK21 program of the Ministry of Education of Korea, 
the CHEP SRC program and Basic Research program 
(grant No.~R01-2005-000-10089-0) of the Korea Science and
Engineering Foundation, and the Pure Basic Research Group 
program of the Korea Research Foundation; 
the Polish State Committee for Scientific Research; 
the Ministry of Education and Science of the Russian
Federation and the Russian Federal Agency for Atomic Energy;
the Slovenian Research Agency;  the Swiss
National Science Foundation; the National Science Council
and the Ministry of Education of Taiwan; and the U.S.
Department of Energy.

\end{document}

%% file: author-conf2007.tex
\affiliation{Budker Institute of Nuclear Physics, Novosibirsk}
\affiliation{Chiba University, Chiba}
\affiliation{University of Cincinnati, Cincinnati, Ohio 45221}
\affiliation{Department of Physics, Fu Jen Catholic University, Taipei}
\affiliation{Justus-Liebig-Universit\"at Gie\ss{}en, Gie\ss{}en}
\affiliation{The Graduate University for Advanced Studies, Hayama}
\affiliation{Gyeongsang National University, Chinju}
\affiliation{Hanyang University, Seoul}
\affiliation{University of Hawaii, Honolulu, Hawaii 96822}
\affiliation{High Energy Accelerator Research Organization (KEK), Tsukuba}
\affiliation{Hiroshima Institute of Technology, Hiroshima}
\affiliation{University of Illinois at Urbana-Champaign, Urbana, Illinois 61801}
\affiliation{Institute of High Energy Physics, Chinese Academy of Sciences, Beijing}
\affiliation{Institute of High Energy Physics, Vienna}
\affiliation{Institute of High Energy Physics, Protvino}
\affiliation{Institute for Theoretical and Experimental Physics, Moscow}
\affiliation{J. Stefan Institute, Ljubljana}
\affiliation{Kanagawa University, Yokohama}
\affiliation{Korea University, Seoul}
\affiliation{Kyoto University, Kyoto}
\affiliation{Kyungpook National University, Taegu}
\affiliation{Ecole Polyt\'ecnique F\'ed\'erale Lausanne, EPFL, Lausanne}
\affiliation{University of Ljubljana, Ljubljana}
\affiliation{University of Maribor, Maribor}
\affiliation{University of Melbourne, School of Physics, Victoria 3010}
\affiliation{Nagoya University, Nagoya}
\affiliation{Nara Women's University, Nara}
\affiliation{National Central University, Chung-li}
\affiliation{National United University, Miao Li}
\affiliation{Department of Physics, National Taiwan University, Taipei}
\affiliation{H. Niewodniczanski Institute of Nuclear Physics, Krakow}
\affiliation{Nippon Dental University, Niigata}
\affiliation{Niigata University, Niigata}
\affiliation{University of Nova Gorica, Nova Gorica}
\affiliation{Osaka City University, Osaka}
\affiliation{Osaka University, Osaka}
\affiliation{Panjab University, Chandigarh}
\affiliation{Peking University, Beijing}
\affiliation{University of Pittsburgh, Pittsburgh, Pennsylvania 15260}
\affiliation{Princeton University, Princeton, New Jersey 08544}
\affiliation{RIKEN BNL Research Center, Upton, New York 11973}
\affiliation{Saga University, Saga}
\affiliation{University of Science and Technology of China, Hefei}
\affiliation{Seoul National University, Seoul}
\affiliation{Shinshu University, Nagano}
\affiliation{Sungkyunkwan University, Suwon}
\affiliation{University of Sydney, Sydney, New South Wales}
\affiliation{Tata Institute of Fundamental Research, Mumbai}
\affiliation{Toho University, Funabashi}
\affiliation{Tohoku Gakuin University, Tagajo}
\affiliation{Tohoku University, Sendai}
\affiliation{Department of Physics, University of Tokyo, Tokyo}
\affiliation{Tokyo Institute of Technology, Tokyo}
\affiliation{Tokyo Metropolitan University, Tokyo}
\affiliation{Tokyo University of Agriculture and Technology, Tokyo}
\affiliation{Toyama National College of Maritime Technology, Toyama}
\affiliation{Virginia Polytechnic Institute and State University, Blacksburg, Virginia 24061}
\affiliation{Yonsei University, Seoul}
  \author{K.~Abe}\affiliation{High Energy Accelerator Research Organization (KEK), Tsukuba} 
  \author{I.~Adachi}\affiliation{High Energy Accelerator Research Organization (KEK), Tsukuba} 
  \author{H.~Aihara}\affiliation{Department of Physics, University of Tokyo, Tokyo} 
  \author{K.~Arinstein}\affiliation{Budker Institute of Nuclear Physics, Novosibirsk} 
  \author{T.~Aso}\affiliation{Toyama National College of Maritime Technology, Toyama} 
  \author{V.~Aulchenko}\affiliation{Budker Institute of Nuclear Physics, Novosibirsk} 
  \author{T.~Aushev}\affiliation{Ecole Polyt\'ecnique F\'ed\'erale Lausanne, EPFL, Lausanne}\affiliation{Institute for Theoretical and Experimental Physics, Moscow} 
  \author{T.~Aziz}\affiliation{Tata Institute of Fundamental Research, Mumbai} 
  \author{S.~Bahinipati}\affiliation{University of Cincinnati, Cincinnati, Ohio 45221} 
  \author{A.~M.~Bakich}\affiliation{University of Sydney, Sydney, New South Wales} 
% ...
\collaboration{The Belle Collaboration}